\documentclass[aps,prd,twocolumn]{revtex4}
\newcommand{\be}{\begin{equation}}
\newcommand{\ee}{\end{equation}}
\newcommand{\bea}{\begin{eqnarray}}
\newcommand{\eea}{\end{eqnarray}}
\newcommand{\bes}{\begin{eqnarray}}
\newcommand{\ees}{\end{eqnarray}}

\begin{document}

%\preprint{}

\title{Inflation via Black Holes with Quantized Area Spectrum}


\author{Tongu\c{c} Rador}

\email[e-mail: ]{rador@gursey.gov.tr}

%\thanks{}

\affiliation{Feza G\"ursey Institute\\ Emek Mah. No:68, 
\c{C}engelk\"oy 81220 \\ Istanbul, Turkey}


\date{November 11, 2002}

\begin{abstract}
I present a very simplistic toy model for the inflationary paradigm where the
size of the universe undergoes a period of exponential growth. The
basic assumption I make use of is that black holes might have a quantized area (mass)
spectrum with a stable ground state and that the universe has started
with a tightly packed collection of these objects alone.

\end{abstract}

\pacs{}
%\keywords{}

\maketitle

\section{Introduction}

Inflationary paradigm \cite{Guth} has been successful in explaining 
away the basic
structural problems of the standard model of cosmology, namely the
horizon, flatness, homogeneity and monopole problems. It is then a
necessary exercise to look for ways it can occur. In what follows 
I present a very naive toy model in
which the scale factor may grow exponentially. I will borrow from
loop quantum gravity where the area operator has been shown to have a
discrete spectrum \cite{Ashtekar} with sizes starting from Planck area. 
The basic physical assumption I will make is that whatever
interaction these objects might have, the outcome of it will have to be
a state present in the spectrum.

\section{The Model and its Application}

It is interesting to think of the possibility of sequences \footnote{I
do not intend to mean that the relation in Eq. (\ref{eq1})
defines the whole spectrum, it is enough for the purposes presented here
that such a subspace can be found and it is shown in \cite{Ashtekar}
that this is possible.} in the
spectrum of black holes such that the mass eigenvalues satisfy,

\be{\label{eq1}}
2 m_{j}=m_{k}\; .
\ee
 
Such a relation is possible as presented in \cite{Ashtekar}. It then follows that two black holes can fuse into one conserving
mass . This interaction will increase the entropy if we assume
Hawking's result \cite{Hawking}  will hold for black holes of any size. The crucial
consequence of this type of interaction is that the initial states have a
radii $r_{j}$ and that the final state has radius $2 r_{j}$. Now let
us assume that we start with $N(0)$ ground state black holes and also
assume that they are tightly packed (say optimum filling for hard spheres). 
Then, it is possible that two neighboring black holes will
fuse into one and that this happens almost simultaneously for the
whole collection. This means that we end up with half the number we
started but the radius of each element grew twice. Let us also assume
that this can be reiterated. Then we find the following,

\bea
N(k)&=&N(0)\; 2^{-k}\;\; ,\\
V_{H}(k)&=&\frac{4 \pi}{3} N(0)r_{0}^{3}\; 2^{2k} \;\; .
\eea

Here $V_{H}$ denotes the total volume behind the horizons, it can be
used to estimate the size of the volume in which the collection is
packed. With the introduction of a filling ratio $\gamma$ the radius of the
collection is given by,

\be
R(k)=\left[(1+\gamma) N(0)\right]^{1/3} r_{0}\; 2^{2k/3} \;\; .
\ee

This represents an exponential growth although the parameter $k$ does
not necessarily have a one-to-one correspondence to the actual time
variable $t$. I will try to estimate the time it takes for a single
step of the mechanism above in the next section.

Now, the initial mass of the collection is about $N(0)$ times the
Planck mass. Assuming that the mass of the universe today is about
$10^{22}$ solar masses \cite{Guth2} we get the following

\be
N(0)\approx 10^{60}\;\; .
\ee

Which will result in,

\be
R(0) \approx 1\;\; {\rm Fermi}.
\ee

This seems large, but let us remember that most of the volume is behind
horizons and that the empty spaces between black holes is
still of the size of Planck length. 

It is clear that this process of halving the number of black holes can
not continue indefinitely. At the extreme case it should stop when
there is only one black hole left. To estimate the final number of black
holes let us assume that the
mechanism stops at a value $k_{f}$ where $R(k_{f})$ becomes about
$1\;\rm cm$. This will give 

\be
k_{f}\approx 65\;\; .
\ee

At this value of $k$ the size of the black holes and hence the inter
spacing between them is of the order of $1$ Fermi. It then starts to
become possible to create proton-antiproton pairs in the inter black
hole space without both of them disappearing behind horizons. It is at
this stage that I assume the mechanism above stops and all the
remaining $N(k_{f})\approx 10^{40}$ black holes explode to release
particles of all sorts (presumably starting with larger mass particles
and, when the inter black hole spacing grows, proceeding to include 
lighter particles) 
such that finally they will settle down to their
ground states \footnote{The actual area spectrum is much richer than the
portion of it I used to climb up the ladder. The black holes
need not use the same steps to get down to their ground states. This
gives more freedom to the evaporation process.}. 
If we borrow that the observed size of the universe
today is about 10 billion light years we get the following relic
density of Planck size black holes

\be
\rho_{{\rm relic}}\approx 10^{-39} \; {\protect{\rm m}}^{-3} \;,
\ee

\noindent that is one relic per a volume of radius about the distance from the
sun to Neptune ; small enough for not having been observed (if they are
evenly distributed in cosmos).

To estimate the temperature we can first eliminate the $t$ dependence
of the extensive variables and get the equation of state

\be{\label{eq:se}}
S^{2}=\kappa\; E V\;\;.
\ee

Here $\kappa$ is a number of order unity times Planck mass squared, (an exact calculation of this
number requires the knowledge of the filling ratio). Then
the temperature is calculated by $T^{-1}=(\partial S/\partial E)_{V}$ after
which we reemphasize the $k$ dependence to get,

\be
T\approx 10^{19}\;2^{-t} \protect{\rm Gev}\;\; .
\ee

Thus the universe starts growing at around the
Planck temperature and the above scenario ends at a temperature of
about $1\;\rm GeV$. The temperature follows the size of the inter
black hole spacing.

\section{Time}

It is evident that a Hamiltonian can not induce transtions between its
own stationay states. So the fusion of the black holes has to be due to
an interaction between them. This interaction will also change the
energy-entropy relation in (\ref{eq:se}). If we new the dynamics we could calculate
the time $\Delta t(k)$ needed for a single step of fusion to occur and
from it we could define a time variable

\be
t(k)=\sum_{n=0}^{n=k} \Delta t(k) \;\; .
\ee

We do not yet know what type of interactions the Plack size black
holes might have. However interestingly enough if we assume a stringy
interaction with a linear potential we get interesting results. Let us
assume that the stringy interaction will  not extend beyond nearest
neigbours since in order to do so the string will have to traverse a
black hole. Thus the contribution of the stringy interaction to the
total energy will go like 

\be
E_{stringy}= \protect{\rm constant}\times N(k) m(k)^{2} r(k) \;.
\ee

Here the constant of proportionality is related to the packing
scheme that is the mean number of nearest neighbours. 
Remembering the $2^{k}$ dependence of all the quantities it is
straightforward to show that the energy
density of the stringy interaction is independent of $k$
and hence of $t$. So like in the inflationary scenario we get  a time
independent and homogeneous \footnote{boundary
effects may change this egergy density slightly} energy density. Having
an interaction potential we can estimate $\Delta t(k)$ for the stringy 
interaction. Newtonian dynamics ($F=ma$) yields 

\be
\Delta t(k) = \protect{\rm constant} \times t_{Pl} \; ,
\ee

which means that $t(k)= {\rm constant} \times k\;\;t_{Pl}$ giving
an exponential growth in ``time'' for the scale factor $R$ of the
universe. 

Another interesting implication is the following. It is not possible
to assume that the stringy interaction dominates when the sizes of the
black holes and hence the mean separation between them grows. Assuming
at some point Newtonian force law sets in the above scheme of
estimation gives,

\be
\Delta t(k) = \protect{\rm constant} \times t_{Pl} \; 2^{k} \; ,
\ee

which results in $R(t)=\protect{\rm constant}\times t^{2/3}$, matter
dominated expansion. It is interesting that this emerges as a byproduct
\footnote {The contribution of the Newtonian interaction to the total energy can
be shown to be a constant (apart from boundary effects) and hence its 
energy density is decaying like $2^{-2k}$ or like $1/t^{2}$.}.
\section{Further Speculations}

The numbers presented depend considerably on the total energy (which
fixes $N(0)$) and on the size of the universe
as it exits the inflationary (with the mechanism presented here)
period which I took to be around $1\; \rm{cm}$. 
As we have seen, it makes sense to identify the exit point
as the point at which the production of particles of mass
corresponding to the mean inter black hole spacing starts. The heaviest known
particles today are the electroweak gauge bosons with a mass of 
around $90\;{\rm
GeV}$. This would correspond to $k_{f}\approx 56$ meaning that
$R(k_{f})\approx 10^{-2}\rm cm$ and $N(k_{f})\approx10^{43}$
resulting in a relic density of about a thousand times larger than
the number I estimated before; one relic in a volume of radius of
roughly twice the distance from the sun to Mars. Much more
massive particles (say a GUT monopole) renders the numbers obtained 
meaningless.
It might however be the case that the fusion process of black holes occurs very
rapidly at the beginning 
such that the heavy particles beyond the scope of standard
model of particle physics may not have enough time to materialize in the
inter black hole regions. 

I assumed trough out that the black holes formed a space and time
independent close packing
scheme. While this is a useful assumption in estimating numbers it is
true that this can not be exactly satisfied. The overall filling ratio may be a
constant but the local filling ratio will fluctuate around a
mean, which would result in density fluctuations; a necessary
ingredient for any cosmological model. On the other hand 
it is also possible that remaining relic
black holes formed local groups and 
eventually duplicated the scenario presented, resulting in fewer
relics, but this somewhat unlikely. Furthermore it could also be the case
that the black holes completely evaporates after the inflationary period
resulting in no relics at all. Finally the model presented here does
not actually need to be the only
mechanism for inflation it might possibly be combined with the usual
inflationary paradigm.

\section{Conclusion}

In this short letter I outlined a very naive  model which
incorporates inflationary paradigm.  It is my hope that the ideas 
presented here have a significance beyond
the toy model presented and may attract the attention of researchers. 
Further analysis requires much more elaborate
mathematical tools. An 
effort to study the significance of the local filling ratio
fluctuations and to incorporate
rigorous interpretation of black hole evaporation is in progress.

I have possibly (not willingly) overlooked relevant literature. I
recently became aware that in loop quantum gravity there is 
a possibility to have a
quantized scale factor for the radius of the universe yielding an
inflationary period \cite{thieman,Bojowald}, the origin of which
however, I believe, is not related to the model presented here.

\begin{acknowledgments}
I thank Christian Hoelbling for mentioning me  Ashtekar's work I
think in around fall 1999. I also thank N.S. De\~ ger for pointing out the
review of T. Thiemann \cite{thieman} on loop quantum gravity from
which I learned of the work of M. Bojowald \cite{Bojowald} on
inflation from loop quantum gravity. 
\end{acknowledgments}


\begin{thebibliography}{10}
\bibitem{Guth} A.H. Guth, Phys. Rev. {\bf D23}, 347, (1981).
\bibitem{Ashtekar} A. Ashtekar, ``Quantum Mechanics of Geometry'',gr-qc/9901023.
\bibitem{Hawking} S. W. Hawking, Commun. Math. Phys. {\bf 43}, 199, (1975).
\bibitem{Guth2} A.H. Guth, ``The Inflationary Universe'', Addison-Wesley,
  (1997), pp. 254.
\bibitem{thieman} T. Thiemann, ``Lectures on Loop Quantum Gravity'', gr-qc/0210094. 
\bibitem{Bojowald} M. Bojowald, ``Inflation from Quantum Gemoetry'', gr-qc/0206054.
\end{thebibliography}

\end{document}